\documentclass[pre,amsmath,amssymb,groupedaddress,twocolumn,showpacs,preprintnumbers, floatfix]{revtex4-1}
\usepackage{graphicx}% Include figure files
\usepackage{dcolumn}% Align table columns on decimal point
\usepackage{bm}% bold math

\begin{document}

\preprint{}

\title{Steplike electric conduction in a classical two-dimensional electron system through a narrow constriction in a microchannel}


% \affiliation command. The \affiliation command should follow the other information
\author{Moto Araki}
\email[]{araki@yukawa.kyoto-u.ac.jp}

\author{Hisao Hayakawa}
%\thanks{}
\affiliation{Yukawa Institute for Theoretical Physics, Kyoto University, Kitashirakawa Oiwake-Cho Sakyo-ku Kyoto 606-8502, Japan}


\date{\today}

\begin{abstract}

Using molecular dynamics simulation, we investigate transport properties of a classical two-dimensional electron system confined in a microchannel with a narrow constriction.
As a function of the confinement strength of the constriction, the calculated conductance in the simulations exhibits steplike increases as reported in a recent experiment [D. G. Rees {\it et al}., Phys. Rev. Lett. \textbf{106}, 026803 (2011)].
It is confirmed that the number of the steps corresponds to the number of stream lines of electrons through the constriction.
We verify that density fluctuation plays a major role in smoothing the steps in the conductance.

\end{abstract}

%73.20.-r 	Electron states at surfaces and interfaces
%73.23.-b 	Electronic transport in mesoscopic systems
%73.20.Qt 	Electron solids 
%02.70.Ns 	Molecular dynamics and particle methods 
%05.10.-a 	Computational methods in statistical physics and nonlinear dynamics
%05.40.-a 	Fluctuation phenomena, random processes, noise, and Brownian motion
%45.50.Jf 	Few- and many-body systems 
%85.30.Hi 	Surface barrier, boundary, and point contact devices 
%73.50.Td 	Noise processes and phenomena 

\maketitle

\section{\label{sec:sec1}Introduction}

Electric conduction in confined geometries has been an important subject of study to characterize properties in the mesoscopic scale \cite{in1}, and its study has a wide possibility of applications to electronics \cite{in2,in3}.
For rarefied gases under one constriction, such as a point contact or a circular orifice, the transport shows the quantized conductance \cite{in14,in15,in1}, Sharvin resistance \cite{in11,in12,in13}, or Maxwell resistance \cite{in16,in17}, depending on the constriction width and the electron mean free path \cite{in4,in5,in6,in7,in8,in9,in10}.
On the other hand, for dense liquids in which the exclusion effects between particles are important, the formation of layers \cite{int4,int5}, pinning and depinning \cite{int6}, and anisotropic and nonuniform mobility \cite{int1,int2,int3} are observed in microchannels.
However, the geometrical effects to conduction in such strongly correlated particle systems have not been fully understood yet.

Recently, Rees {\it et al}. have found the existence of steplike conductance in the conduction of classical two-dimensional (2D) electrons on liquid ${}^4$He (Refs. \cite{i1} and \cite{i2}) using a device with point-contact geometry \cite{Rees}.
They suggest that the origins of the conductance are not quantum effects, as in the conductance quantization \cite{in14}, but are the effects of strong correlations in classical electron systems.
We find it necessary to justify their suggestion and to clarify the mechanisms of the dynamics behind the steplike conduction of electrons on liquid ${}^4$He.
The main goal of this paper, thus, is to understand the mechanisms of the steplike conduction in classical 2D electron systems.

In this paper, we investigate the electric conduction of the classical 2D electron system through a narrow constriction in a microchannel using a molecular dynamics (MD) simulation with two particle baths under a No\'se-Hoover thermostat.
The electrons in the MD simulations are confined in a point-contact-like shape and interact with each other, in terms of electric potential derived from the Poisson equation under a boundary condition given so as to imitate the device in Ref. \cite{Rees} (see Fig. \ref{f1}).
Our model and method stand on the presumption that the many-body effects in the confined geometry are essential for the steplike electric conduction (see Sec. \ref{sec:subsec2-0} and Appendix \ref{sec:a1}).
We calculate conductance as a function of the confinement strength of the constriction.
To confirm the suggestion by Rees {\it et al}. \cite{Rees} and to develop the understanding of the steplike conduction, we investigate the static and dynamical properties of electrons near the constriction from the spatial distribution of electron density, electrostatic potential, and potential fluctuation.

The organization of the paper is as follows.
In Sec. \ref{sec:sec2}, our model, methods, and simulation details are provided.
The conductance and the other quantities calculated in our simulations are presented in Sec. \ref{sec:sec3}, in which the suggestion by Rees {\it et al}. \cite{Rees} is also verified.
Based on the results in Sec. \ref{sec:sec3}, we discuss the mechanisms causing the observed conductance in Sec. \ref{sec:subsec41}.
In Sec. \ref{sec:subsec42}, we compare our results with the observed conductance in the experiment.
We also present the result of the Langevin dynamics (LD) simulation method which is the molecular dynamics using the Langevin thermostat in Sec. \ref{sec:subsec43}.
The results and the discussions are summarized in Sec. \ref{sec:sec5}.
In appendices, we present some detailed descriptions of our model and method.
In Appendix \ref{sec:a1}, we estimate the resistance from the electron-helium vapor atom scattering and from the electron-ripplon scattering.
In Appendix \ref{sec:a2}, we give the details of the method to calculate physical quantities.
In Appendix \ref{sec:a3}, we briefly explain the method of our LD simulation.

\section{\label{sec:sec2}Model, and Numerical Methods}

In this section, we introduce our model with a demonstration to justify our treatment for the electrons in the device in Ref. \cite{Rees} and explain the method of the MD, including how to provide the simulation setup and to calculate observed quantities.

\subsection{\label{sec:subsec2-0}Electron states and transport properties in the electron system over liquid ${}^4$He on metal electrodes}

In this section, we briefly explain the electron state and the transport process of electrons in the device in Ref. \cite{Rees}.
The system of electrons over liquid ${}^4$He on metal electrodes studied by Rees {\it et al}. \cite{Rees} is an ideal system to investigate strongly correlated classical 2D electron systems \cite{int3,i1,i2,i12,i22,Rees}.
The electrons are confined on a plane at the height $z_{\rm{G}}$ from the liquid ${}^4$He due to a potential barrier at the liquid surface, an interaction between the electrons and polarized liquid ${}^4$He, and charges induced in the metal electrodes \cite{i24}.
We also apply a holding electric field $E_{\perp}$ normal to the liquid surface.
For surface density $n_{\rm{s}}\sim 10^{8}-10^{9}$cm$^{-2}$, temperature $T\sim 1$ K, and liquid thickness $z_{\rm{He}}\sim 1 \ \mu$m, the vertical motion of each electron can be decoupled with that in the parallel direction to the surface.
For the vertical direction, the electrons occupy the ground state of the quantized electron states, whereas the parallel motion is classical.
These treatments of electron motion on the plane can be justified because the interelectron separation, $r_{s}=2/\sqrt{\pi n_{\rm{s}}}\sim 10^{-1} \ \mu$m, is much larger than the thermal de Broglie wavelength, $\lambda_{D}\sim 10^{-2} \ \mu$m, and the energy gap between the ground and the first excited states, $\Delta \sim 19$ K, is sufficiently larger than the temperature in the experiment.
Here, $\Delta$ is estimated by the variational method \cite{i25} under $E_{\perp}=6.5\times 10^{2}$ V$/$cm in a very similar device \cite{i22} to that used in Ref. \cite{Rees}.

For the bulk 2D electrons, it is known that the mobility on the surface is little affected by the scatterings by roughness on the interface liquid substrate for $z_{\rm{He}}\sim 1 \ \mu$m \cite{i28-0,i28,i28-2,i29}, but is dominated by the scatterings between electrons and helium gas atoms for $T > 1$ K or between electrons and ripplons at lower temperature \cite{i24,i27,int3,i1}.
The correlation effects between electrons on the transport have been discussed by the kinetic equation method under the complete control approximation \cite{cc1,i25-2,cc2,cc3,cc4}, the force-balance method \cite{fb1,fb2,i29}, and more sophisticated theory \cite{i1,An,DF1}.

In Appendix \ref{sec:a1}, we briefly estimate the resistance for the transport in the channel geometry as in Ref. \cite{Rees}.
The result supports that the many-body effects in the confined geometry are dominant for the steplike electric conduction.

\subsection{\label{sec:subsec2-1}Interaction and confining potentials}
%%%%%%%%%%%%%%%%%
\begin{figure}[bh!]
\center\includegraphics[width=60mm,clip]{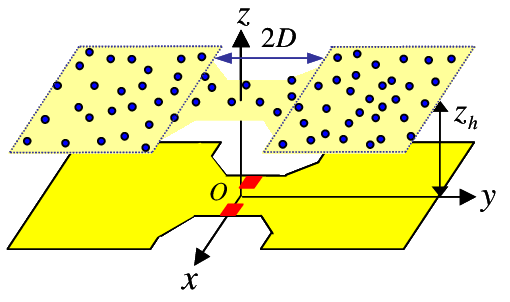}
\caption{(Color online) A schematic view of our simulation setup of a classical 2D electron system confined on the plane at $z=z_{h}$. On the boundary plane $z=0$, the yellow and red regions, respectively, correspond to the reservoir and the split-gate electrodes in the experiment \cite{Rees}.
The system consists of a left and a right reservoirs, and a channel of length $2D$, where the reservoirs are connected with the channel.
}
\label{f1}
\end{figure}
%%%%%%%%%%%%%%%%%

In this section, we present the forms of interaction potential between electrons and confining potential for our MD.
The electric potential energy for the $i-$th electron on liquid $^{4}$He is given by \cite{r2-1}
\begin{eqnarray}
\phi(\mathbf{r}_{i},t)=\sum_{j ( \neq i )}\phi_{\rm{I}}\left[r_{ij}(t)\right]+\phi_{\rm{C}}\left[\mathbf{r}_{i}(t)\right],
\label{e0}
\end{eqnarray}
where  $\mathbf{r}_{i}(t)=\left[ x_{i}(t),y_{i}(t) \right]$ is the position of the $i-$th electron at $z=z_{h}$ at time $t$, $r_{ij}(t)\equiv \vert\mathbf{r}_{i}(t)-\mathbf{r}_{j}(t)\vert$, $\phi_{\rm{I}}(r)$ is the interaction potential energy between two surface electrons, and $\phi_{\rm{C}}(\mathbf{r})$ is the confining potential energy.
We solve the 3D Poisson equation for $\phi$ in the semi-infinite domain $z\geq 0$, and presume the electrons to be confined at the height $z = z_{h}$ from the plane $z = 0$.
Here, we set $z_h$ as the thickness of liquid $^{4}$He, $z_{h}=1.5 \ \mu$m, which follows the experimental setup \cite{Rees} (Fig. \ref{f1}), where the average distance $z_{\rm{G}}\sim 10^{-2} \ \mu$m of the hovering electrons from the liquid surface is disregarded \cite{i25}.
Therefore, the potential is obtained from an analytic solution at $z = z_{h}$ of the Poisson equation.
Then, $\phi_{\rm{I}}(r)$ is given by \cite{r2-2,r2-3}
\begin{equation}
\phi_{\rm{I}}(r)=
e^{2}\left[ \frac{1}{r}-\frac{1}{\sqrt{r^2+4z_{h}^{2}}}\right],
\label{e1}
\end{equation}
where $e$ is the elementary electric charge.
The right-hand side in Eq. (\ref{e1}) consists of the bare Coulomb interaction and the dominant screening effects between surface electrons which represents the contribution of the image charge induced in the metal electrodes.
Equation (\ref{e1}) is obtained under the same order approximation to the potential energy in the previous studies \cite{i24,i25}.

\begin{figure}[bh!]
\center\includegraphics[width=86mm,clip]{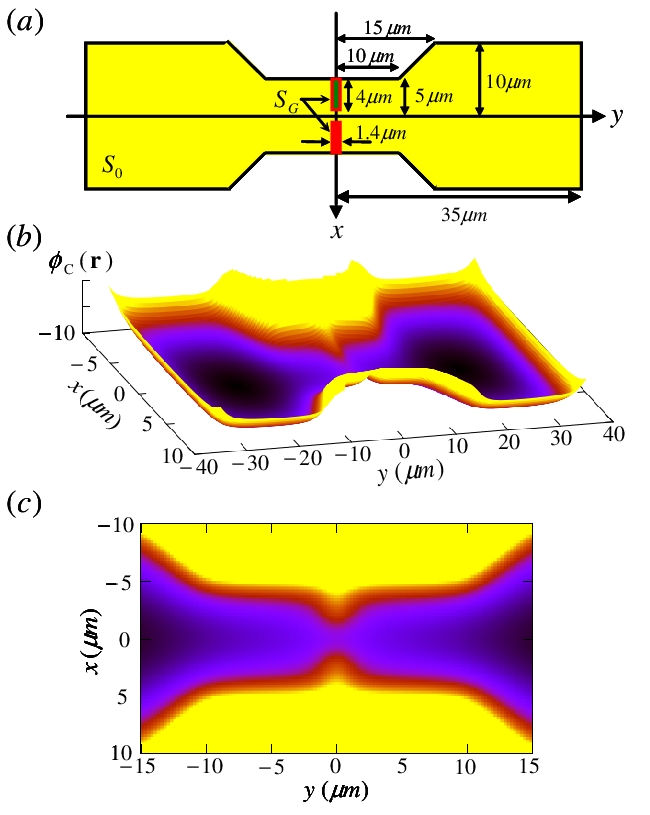}
\caption{(Color online) (a) The boundary condition at $z=0$: $S_{0}$ and $S_{\rm{G}}$, respectively, are the yellow and red regions. The electric potential in $S_{0}$ and the outside of the colored regions are $V_{0}$ and $0$, respectively.
The gate voltage $V_{\rm{G}}$ is imposed to $S_{\rm{G}}$.
(b) The confining potential energy $\phi_{\rm{C}}$ for $V_{0}=0.38$ V, $V_{\rm{G}}=0.38$ V, and $z_{h}=1.5 \ \mu$m in Eq. (\ref{e2}).
(c) The contour plot of $\phi_{\rm{C}}$ in the channel, which is the region of $\vert y \vert < D=15\ \mu$m.}
\label{f2}
\end{figure}

The boundary conditions we impose on the Poisson equation are 
\begin{eqnarray}
\left\{%
\begin{array}{llll}
\phi(\mathbf{r},z=0) & = & V_{0}      & (\mathbf{r} \in S_{0})      \\
\phi(\mathbf{r},z=0) & = & V_{\rm{G}} & (\mathbf{r} \in S_{\rm{G}}) \\
\phi(\mathbf{r},z=0) & = & 0          & (\mathbf{r} \notin S_{0}\cup S_{\rm{G}}) \\
\phi(\mathbf{r},z)   & = & 0          & (\vert \mathbf{r} \vert\to\infty) \\
\phi(\mathbf{r},z)   & = & 0          & (z\to \infty)
\end{array}
\right. \nonumber
\end{eqnarray}
where $S_{0}$, $S_{\rm{G}}$, and the outside of $S_{0}\cup S_{\rm{G}}$ at $z=0$, respectively, represent the reservoir, the split gate, and the guard electrodes \cite{Rees} [see Fig. \ref{f2}(a)].
Then, $\phi_{\rm{C}}(\mathbf{r})$ is represented as
\begin{eqnarray}
\phi_{\rm{C}}(\mathbf{r})&=&-\frac{eV_{0}}{4 \pi}\int_{S_{0}}dx_{0}dy_{0}\left[ \frac{2z_{h}}{[\vert\mathbf{r}-\mathbf{r}_{0}\vert^2+z_{h}^{2}]^{3/2}}\right]
\nonumber \\
& &-\frac{eV_{\rm{G}}}{4 \pi}\int_{S_{\rm{G}}}dx_{0}dy_{0}\left[ \frac{2z_{h}}{[\vert\mathbf{r}-\mathbf{r}_{0}\vert^2+z_{h}^{2}]^{3/2}}\right],
\label{e2}
\end{eqnarray}
where $V_{0}$ and $V_{\rm{G}}$ are the voltages of a reservoir and a split-gate, respectively.
It should be noted that each integration in Eq. (\ref{e2}) can be performed exactly if we assume that the integration range consists of rectangles and triangles, as in Fig. \ref{f2}(a).

We also adopt $V_{0}=0.38 $ V and $V_{\rm{G}}$ of the range from $-0.05$ V to $0.38$ V, which imitates the device in Ref. \cite{Rees}.
When $V_{\rm{G}}$ is set to $V_{0}$, $\phi_{\rm{C}}$ works as the confinement without the point contact, and when $V_{\rm{G}}$ is set to a voltage lower than $V_{0}$, $\phi_{\rm{C}}$ additionally generates the point contact due to a voltage induced between $S_{0}$ and $S_{\rm{G}}$, as seen in Figs. \ref{f2}(b) and (c).

%%%%%%%%%%%%%%%%%
%%%%%%%%%%%%%%%%%

\subsection{\label{sec:subsec2-2}Constant temperature and chemical potential molecular dynamics}
Our MD is built on a hybrid scheme of a constant temperature MD with a No\'se-Hoover thermostat \cite{s1,s2,s3} and a constant chemical potential MD (CMD) \cite{s4,s5,s6,s7}.
The reason we adopt the No\'se-Hoover thermostat is as follows: This method is the established one to reproduce the precise equilibrium state, and the equation of motion for electrons can keep the local time-reversal symmetry.
In CMD, we introduce a fractional particle characterized by an extended number variable (ENV) for a particle bath, where the integer part of the ENV denotes the number of particles, and the fractional part represents the value of the fractional particle.
The ENV couples with the system through the fractional particle interacting with the rest of the system.

In our simulation, the system consists of a left and a right reservoir, and a channel of length $2D$ with $D=15 \ \mu$m, where the reservoirs are connected with the channel (see Figs. \ref{f1} and \ref{f2}).
The left (right) reservoir has an electrochemical potential $\mu_{\rm{L}}$ ($\mu_{\rm{R}}$) which can be divided into two parts as,
\begin{eqnarray}
\mu_{\gamma}=\mu^{\rm{0}}_{\gamma}+\mu^{\rm{1}}_{\gamma},
\label{ecp}
\end{eqnarray}
with $\gamma=\rm{L},\rm{R}$, where $\mu^{0}_{\gamma}$ is the intrinsic part of the ideal chemical potential \cite{HM}, and  $\mu^{\rm{1}}_{\gamma}$, which is the control parameter in our MD, is the sum of the excess chemical potential and the confining potential energy.
The left (right) reservoir consists of one fractional particle and temporally variational $N_{\rm{L}}$ ($N_{\rm{R}}$) electrons in $S_{\rm{L}}$ ($S_{\rm{R}}$), where $S_{\rm{L}}$ ($S_{\rm{R}}$) is the region $y<-H$ ($y>H$) with $H=15 \ \mu$m.

The equation of motion for the electrons in our MD is given by
\begin{eqnarray}
m\frac{d^{2}r_{i}^{\alpha}}{dt^{2}}
&=&
-\frac{\partial\phi (\mathbf{r}_{i})}{\partial r_{i}^{\alpha}}
-m\dot{\zeta}\dot{r_{i}}^{\alpha}
\nonumber \\
& &-\theta(-y_{i}-H)\left[\nu_{\rm{L}}-N_{\rm{L}}\right]\frac{\partial\phi_{\rm{I}} (r_{i\rm{L}})}{\partial r_{i}^{\alpha}}
\nonumber \\
& &-\theta(y_{i}-H)\left[\nu_{\rm{R}}-N_{\rm{R}}\right]\frac{\partial\phi_{\rm{I}} (r_{i\rm{R}})}{\partial r_{i}^{\alpha}} ,
\label{e3-1}
\end{eqnarray}
where $r_{i}^{\alpha}$ and $r_{\rm{L}(\rm{R})}^{\alpha}$ are, respectively, the $\alpha$ componenti$\alpha =x,y$jof the position of the $i-$th electron and the fractional particle belonging to the left (right) reservoir, $r_{i\rm{L}(\rm{R})}\equiv \vert\mathbf{r}_{i}-\mathbf{r}_{\rm{L}(\rm{R})}\vert$, $\nu_{\rm{L}}$ ($\nu_{\rm{R}}$) is the ENV of the left (right) reservoir, $m$ is the electron mass, and $\theta(x)$ is a step function, i.e., $\theta(x)=1$ for $x>1$, and $\theta(x)=0$ otherwise.
Here, the ``friction" coefficient $\zeta$ is adjusted according to the following equation \cite{s3}:
\begin{eqnarray}
Q_{\zeta}\frac{d^{2}\zeta}{dt^{2}}&=&2\left[ \sum_{i}\frac{m\dot{r}_{i}^{2}}{2}-Nk_{B}T_{\rm{K}} \right] ,
\label{e3-2}
\end{eqnarray}
where $Q_{\zeta}$ is the ``mass parameter" of $\zeta$, $N$ is the temporally variational total number of electrons, $k_{B}$ is the Boltzmann constant, and $T_{\rm{K}}$ is the expected kinetic temperature.  
The time evolutions of the fractional particle coordinate $r_{\gamma}^{\alpha}$ and the ENV $\nu_{\gamma}$ are respectively given by
\begin{eqnarray}
m\frac{d^{2}r_{\gamma}^{\alpha}}{dt^{2}}
=
-(\nu_{\gamma}-N_{\gamma})\left[\sum_{i\subset S_{\gamma}}\frac{\partial\phi_{\rm{I}} (r_{\gamma i})}{\partial r_{\gamma}^{\alpha}}+\frac{\partial\phi_{\rm{C}} (\mathbf{r}_{\gamma})}{\partial r_{\gamma}^{\alpha}}\right],
\label{e3-3}
\end{eqnarray}
\begin{eqnarray}
Q_{\nu}\frac{d^{2}\nu_{\gamma}}{dt^{2}}
=
\mu_{\gamma}^{1}-\left[\sum_{i\subset S_{\gamma}}\phi_{\rm{I}}(r_{\gamma i})+\phi_{\rm{C}}(\mathbf{r}_{\gamma})\right] ,
\label{e3-5}
\end{eqnarray}
where $Q_{\nu}$ is the mass parameter of $\nu_{\gamma}$.

The temporal variations of $N_{\rm{L}}$ and $N_{\rm{R}}$ are governed by the following protocol \cite{s7}.
When $\nu_{\gamma}-N_{\gamma}$ becomes zero, we delete the fractional particle, and replace one electron in the reservoir by a new fractional particle satisfying $\ddot{\nu}_{\gamma}^{\rm old}=\ddot{\nu}_{\gamma}^{\rm new}$, when the coordinate and the velocity for the deleted fractional particle are discarded, and those of the replaced electron are given over to the new fractional particle.
When $\nu_{\gamma}-N_{\gamma}$ becomes one, we convert the fractional particle into an electron, and insert a new fractional particle at the position where the fractional particle satisfies $\ddot{\nu}_{\gamma}^{\rm old}=\ddot{\nu}_{\gamma}^{\rm new}$ and the potential energy also satisfies a local minimum condition $\ddot{\mathbf{r}}_{\gamma}\simeq 0$, when the coordinate and the velocity of the converted fractional particle are given over to the new electron, and the velocity of the new fractional particle is set to zero.
The condition $\ddot{\nu}_{\gamma}^{\rm old}=\ddot{\nu}_{\gamma}^{\rm new}$ ensures the temporal continuity of $\ddot{\nu}_{\gamma}$, and $\ddot{\mathbf{r}}_{\gamma}\simeq 0$ works so as not to change the average velocity of the system due to the insertions of a fractional particle.
Moreover, the inserting place of the fractional particle in the left (right) reservoir is selected from the squares created by dividing area $\rm{A}_{\rm{L}}$ ($\rm{A}_{\rm{R}}$) in  $S_{\rm{L}}$ ($S_{\rm{R}}$) into $0.01 \ \mu$m square mesh, based on the condition described above.
The behavior of electrons in the channel is not disturbed by the fractional particles, because if the fractional particles try to enter the channel, $\ddot{\nu}_{\gamma}$ increases, and thus the fractional particle is converted into an electron according to the above protocol for the particle conversion.

\subsection{\label{sec:subsec2-3}Simulation setup}
Throughout our MD simulations the initial total number of electrons $N_{\rm{I}}$ is $1284$, and the initial positions of electrons are located in the ground-state configuration which forms a classical 2D Wigner crystal deformed by the confinement at each $V_{\rm{G}}$.
The initial velocities of electrons are randomly assigned from the Maxwell-Boltzmann distribution at $T_{\rm{K}}=1.2$ K corresponding to the experimental setup \cite{Rees}.
The initial positions of the fractional particles are determined based on the protocol in the case of $\ddot{\nu}_{\gamma}=0$, and their initial velocities are set to zero, $\dot{\mathbf{r}}_{\gamma}= 0$.
The extension variables are initially set to: $\nu_{\gamma}=N_{\gamma}+0.5$, $\dot{\nu}_{\gamma}=0$, $\zeta=0$, and $\dot{\zeta}=0$.
We choose $Q_{\zeta}$ as: $Q_{\zeta}=k_{B}T_{\rm{K}}N_{\rm{I}}\tau_{\rm{N}}^{2}=6.02\times 10^{-20}$ meV$\cdot$s$^{2}$, where $\tau_{\rm{N}}=2.13\times 10^{-11}$ is the No\'se-Hoover relaxation time.
The value of $\tau_{\rm{N}}$ is to make the thermostat work effectively \cite{s8}, and is close to the characteristic period of short-wavelength vibrations of the bulk 2D Wigner crystal \cite{s10}, $\tau_{\rm{s}}=8.52\times 10^{-11}$ s.
It is believed that the temporal variation of temperature does not affect long-wavelength conductivity directly because the temperature fluctuation induces only short-wavelength fluctuations in the momentum of electrons.
We also carry out the MD simulations with another mass parameter $Q'_{\zeta}=16Q_{\zeta}$, and then obtain quantitatively similar results for all quantities calculated in the following.
Although it is known that $Q_{\nu}$ affects the cycle of fluctuation in the number of particles, 
we only use one parameter $Q_{\nu}=E_{0}t_{0}^2$, where $E_{0}=1.44$ meV and $t_{0}=6.28\times 10^{-11} $ s are our MD units of energy and time, respectively.
Because of the restriction of our computer resources, we have to limit the area $\rm{A}_{\rm{L}}$ ($\rm{A}_{\rm{R}}$) to being the region $-y_{\rm{max}}<y<-y_{\rm{min}}$ ($y_{\rm{min}}<y<y_{\rm{max}}$) with $y_{\rm{min}}=31.7 \ \mu$m and $y_{\rm{max}}=32.2 \ \mu$m in $\vert x\vert < 7 \ \mu$m, where $-y_{\rm{min}}$ ($y_{\rm{min}}$) corresponds to the average position of the electron present at the furthest left (right) in all of the electrons in the left (right) reservoir.
The above initial conditions and the parameters are used throughout our simulations.

We fix $\mu^{1}_{\rm{L}}$ and $\mu^{1}_{\rm{R}}$ as $\mu_{\rm{L}}^{1}=49.16$ meV and $\mu_{\rm{R}}^{1}=48.96$ meV, where we set the minimum of the confining potential energy to zero.
Thus the chemical-potential difference $\Delta \mu^{1}\equiv \mu^{1}_{\rm{L}}-\mu^{1}_{\rm{R}}$ induces a direct current (dc) between the reservoirs.
Although Rees {\it et al}. \cite{Rees} measured an alternative current (ac) conductance, we have measured dc conductance, because the cycles applied in the experiment \cite{Rees}, $5.0\times 10^{-6} $ s, corresponding to $10^{7}$ steps of our MD are too long.
We also carry out the MD simulations with $\Delta \mu^{1}=0.1$, $0.4$, $1.0$, and $2.0$ meV, and then obtain much the same conductance for $\Delta \mu^{1}=0.1$ and $0.4$ meV as $\Delta \mu^{1}=0.2$ meV, and confirm no steplike structure in conductance for $\Delta \mu^{1}=1.0$ and $2.0$ meV  (see Supplemental Material) \cite{SM}.
For the $\mu^{1}_{\gamma}$ we set, the averaged density $\bar{n}$ over the channel is $2.32(\pm 0.02) \times 10^{8}$ cm$^{-2}$, where we enclose the standard deviation for variation with $V_{\rm{G}}$ in the parentheses, and use its notation throughout this paper.
Thereby, the 2D electrons belong to a liquid state for the bulk system \cite{mel1} because the plasma parameter $\Gamma = e^{2}\sqrt{\pi \bar{n} }/k_{\rm{B}}T_{\rm{K}}=37.6$ is much smaller than the critical plasma parameter \cite{mel2} $\Gamma_{\rm{c}} = 137$.

\subsection{\label{sec:subsec2-4}Conductance calculation}
Under the above setup, we calculate the electric current, density distribution, potential energy, fluctuation of potential energy, etc. in a steady state after 500,000 steps from the initial state (see Appendix \ref{sec:a2} for details).
In order to obtain the conductance
\begin{eqnarray}
G(V_{\rm{G}})=-eI(V_{\rm{G}})/\Delta\mu ,
\label{conductance}
\end{eqnarray}
as a function of $V_{\rm{G}}$, we compute the electric current in the $y$ direction in the channel at each $V_{\rm{G}}$,
\begin{eqnarray}
I(V_{\rm{G}})=-e\frac{\left\langle\sum_{i} \theta \left[ D-\vert y_{i}(t)\vert \right] \dot{y}_{i}(t) \right\rangle }{2D},
\label{current}
\end{eqnarray}
under the assumption $\Delta\mu=\Delta\mu^{1}$.
In this paper, $\langle \cdots \rangle$ represents both the ensemble and the time averages, where the ensemble is generated with the different random seeds for the initial velocity distribution.
Through the current calculation, it is confirmed that the replacement of $D$ in Eq. (\ref{current}) by $y_{\rm{min}}$ or $y_{\rm{max}}$ lowers $I$ at each $V_{\rm{G}}$ only by $1$\% or $2$\%.
The contribution of $\Delta\mu^{0}$ to $G$ is negligible because the corresponding quantity
\begin{eqnarray}
\Delta \hat{\mu}^{0}=\left\{ \int_{y<0}-\int_{y>0} \right\} d^2\mathbf{r} \ f(\mathbf{r})k_{\rm{B}}T_{\rm{K}}\ln \left[ \lambda_{\rm{D}}^{2} n(\mathbf{r}) \right],
\label{C3}
\end{eqnarray}
is of the order of $10^{-3}\Delta\mu^{1}$.
We also calculate $G$ in the system with the reservoirs widened from the $20 \times 20 \ \mu$m$^{2}$ squares to the $20\times 25 \ \mu$m$^{2}$ rectangles.
$G$ in the widened system is in the range of the error bar of $G$ in the original system all over $V_{\rm{G}}$.
Moreover, we calculate $G$ of the widened system with $H=20 \ \mu$m, in which the obtained $G$ in the widened system is in agreement with $G$ in the original system.

\section{\label{sec:sec3}Results}

%%%%%%%%%%%%%%%%%%%%%%%%%%%%%%%%%%
\begin{figure}[bh!]
\center\includegraphics[width=86mm,clip]{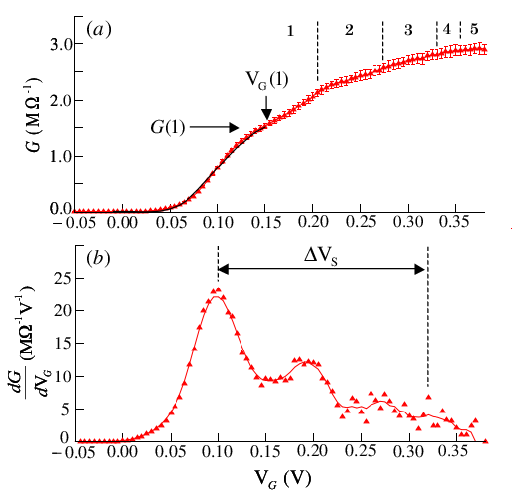}

\caption{(Color online) (a) The dc conductance $G$ obtained from the MD simulations versus the gate voltage $V_{\rm{G}}$, where each error bar represents the standard deviation of the ensemble average for each $G$.
Here, $V_{\rm{G}}(1)=0.15$ V is the gate voltage at which the potential barrier $E_{\rm{B}}$ disappears, and $G(1)$ is $G$ at $V_{\rm{G}}(1)$.
The solid line is the approximate conductance $\tilde{G}$ in Eq. (\ref{e5}).
The vertical dashed lines and the number between them indicate the number of the stream lines organized by electrons in the gate, observed directly in the density distribution $n(\mathbf{r})$.
(b) $dG/dV_{\rm{G}}$ with respect to $V_{\rm{G}}$ (red triangles) and the derivative of five-point unweighted smoothed $G$ (red solid line).
Here, the range $\Delta V_{\rm{S}}= 0.202$ V between the vertical dashed lines runs from the first to the fourth peak in $dG/dV_{\rm{G}}$.
}
\label{fr1}
\end{figure}
%%%%%%%%%%%%%%%%%%%%%%%%%%%%%%%%%%

In this section, we present the obtained results from our simulations.
The results clarify the origin of the steps \cite{Rees}, and give clues for understanding the electron dynamics behind the steplike conduction.

In Fig. \ref{fr1}(a), we plot $G$ under the time average over $4,300,000$ time steps and the ensemble average over $170$ different initial conditions, where $G$ exhibits weak steplike increases.
Moreover, $dG/dV_{\rm{G}}$ in Fig. \ref{fr1}(b) shows a characteristic oscillation similar to that in Ref. \cite{Rees}.

%%%%%%%%%%%%%%%%%%%%%%%%%%%%%%%%%%
\begin{figure}[!t]
\center\includegraphics[width=86mm,clip]{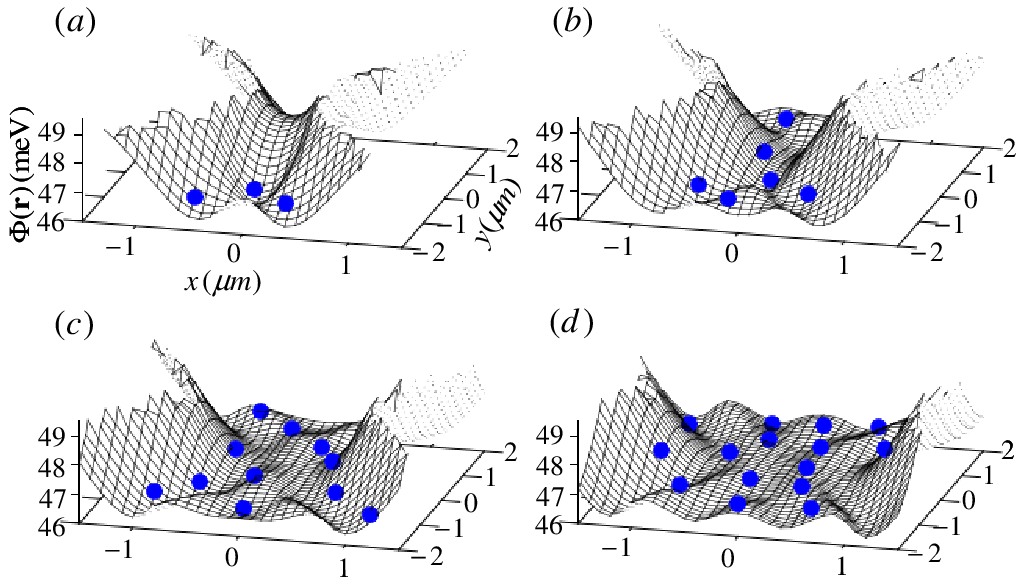}
\caption{(Color online) The spatial distribution of the effective potential energy $\Phi(\mathbf{r})$ near the gate at $V_{\rm{G}}=$ (a) $0.09$ V, (b) $V_{\rm{G}}=0.18$ V, (c) $V_{\rm{G}}=0.25$ V, and (d) $V_{\rm{G}}=0.31$ V.
Here, the blue spheres are placed at an electron configuration of a step in our simulations.
}
\label{f7}
\end{figure}
%%%%%%%%%%%%%%%%%%%%%%%%%%%%%%%%%%

The insulation for low $V_{\rm{G}}$ is due to the existence of a potential barrier in the gate; on the other hand, the increase in $G$ for high $V_{\rm{G}}$ is for expansion of the width of a constriction in the gate.
Figure \ref{f7} depicts the spatial distribution of the effective potential energy $\Phi(\mathbf{r})$ near the gate [see Eq. (\ref{ep})].
We should keep in mind that $\Phi(\mathbf{r})$ is the sum of the confining potential and averaged electrostatic potential from self-organizing distributed electrons.
As seen in Fig. \ref{f7}(a), the existence of the high-energy barrier can be verified for low $V_{\rm{G}}$.
Figures \ref{f7}(b)-(d) also display the decrease of the barrier height and the increase in the constriction width as $V_{\rm{G}}$ increases.

%%%%%%%%%%%%%%%%%%%%%%%%%%%%%%%%%%
\begin{figure}[!t]
\center\includegraphics[width=86mm,clip]{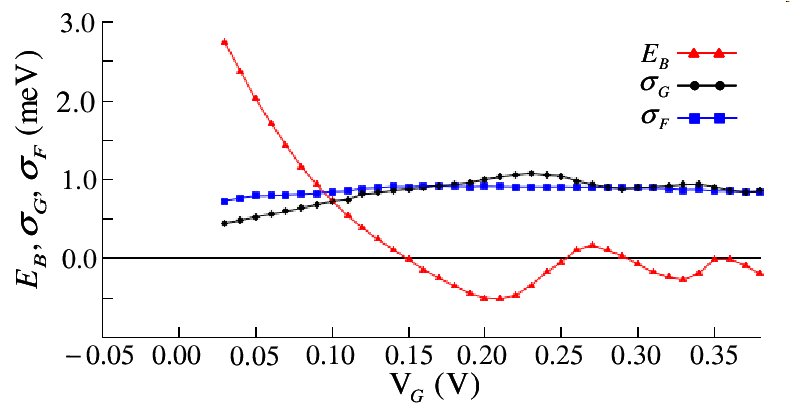}
\caption{(Color online) The potential barrier $E_{\rm{B}}$ in Eq. (\ref{e4}) (red triangles), and the standard deviations of temporally variational potential energy at the positions of the center of the gate, $\sigma_{\rm{G}}$ (black circles), and in front of the gate, $\sigma_{\rm{F}}$ (blue rectangles).
}
\label{fr2}
\end{figure}
%%%%%%%%%%%%%%%%%%%%%%%%%%%%%%%%%%

With the aid of $\Phi(\mathbf{r})$, we introduce the energy barrier defined by
\begin{eqnarray}
E_{\rm{B}}(V_{\rm{G}})&\equiv &
\Phi (\mathbf{r}_{\rm{G}})- \int_{S_{\rm{F}}} d^2\mathbf{r} f(\mathbf{r};S_{\rm{F}})\Phi (\mathbf{r}),
\label{e4}
\end{eqnarray}
which represents the potential energy difference between the center position in the gate, $\mathbf{r}_{\rm{G}}=(0,0)$, and the region in front of the gate, $S_{\rm{F}}=\{(x,y) \vert y_{\rm{F}}\leq y\leq 0 \}$, with $y_{\rm{F}}=-1.7 \ \mu$m.
Here, $y_{\rm{F}}$ is selected so as to include an average position of the electron just in front of the gate, and $f(\mathbf{r};S_{\rm{F}})=n(\mathbf{r})/\bar{N}(S_{\rm{F}})$ with $\bar{N}(S_{\rm{F}})=\int_{S_{\rm{F}}} d^2\mathbf{r}n(\mathbf{r})$ is the normalized single-particle distribution function in $S_{\rm{F}}$.
Figure \ref{fr2} shows that $E_{\rm{B}}$ becomes zero at $V_{\rm{G}}=0.15$ V ($\equiv V_{\rm{G}}(1)$) and the barrier exists for $V_{\rm{G}}<V_{\rm{G}}(1)$.
However, a tiny current exists at $V_{\rm{G}}=0.05 \ $V in Fig. \ref{fr1}(a), although the electrons cannot get over the barrier ($E_{\rm{B}}=2.02$ meV) because of the small kinetic energy $k_{B}T_{\rm{K}}=0.103 \ $ meV at $T_{\rm{K}}=1.2 \ $K.
The origin of the current will be discussed in Sec. \ref{sec:subsec41}.

%%%%%%%%%%%%%%%%%%%%%%%%%%%%%%%%%%
\begin{figure}[!t]
\center\includegraphics[width=86mm,clip]{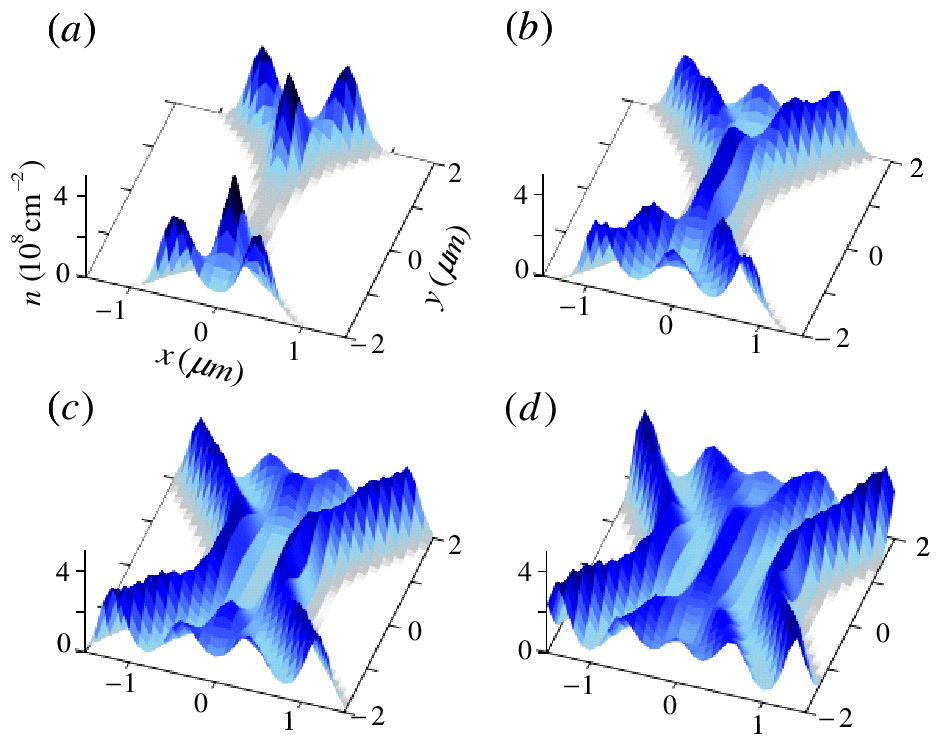}
\caption{(Color online) The density distribution $n(\mathbf{r})$ near the gate at (a) $V_{\rm{G}}=0.05$ V, (b) $V_{\rm{G}}=0.15$ V, (c) $V_{\rm{G}}=0.24$ V and (d) $V_{\rm{G}}=0.30$ V.
The lines observed in (b), (c), and (d) are formed by sequential single-electron flow.
}
\label{f6}
\end{figure}
%%%%%%%%%%%%%%%%%%%%%%%%%%%%%%%%%%

The steplike behavior in $G$ is roughly understood by examining the density distribution function $n(\mathbf{r})$ [see Eq. (\ref{n})].
Figure \ref{f6} illustrates typical density patterns of electrons near the gate, corresponding to the $s-$th step of $G$.
For low $V_{\rm{G}}$, as seen in Fig. \ref{f6}(a), the electrons seem to stay in front of the gate but flow slightly through the barrier.
We can directly observe that the electrons flow through the gate in one line at the first step of $G$ [Fig. \ref{f6}(b)]), two lines at the second step [Fig. \ref{f6}(c)] and three lines at the third step [Fig. \ref{f6}(d)].
Therefore, it is confirmed that the steplike increases are not originated from the conductance quantization \cite{in14}, but the effects can be attributed to the increment in the number of stream lines of electron flow in the constriction, as suggested by Rees {\it et al}. \cite{Rees}.
However, the smooth steps in $G$ cannot be understood only with the discrete increments in the number $N_{\rm{G}}$ of electrons to pass simultaneously through the gate.

%%%%%%%%%%%%%%%%%%%%%%%%%%%%%%%%%%
\begin{figure}[!t]
\center\includegraphics[width=86mm,clip]{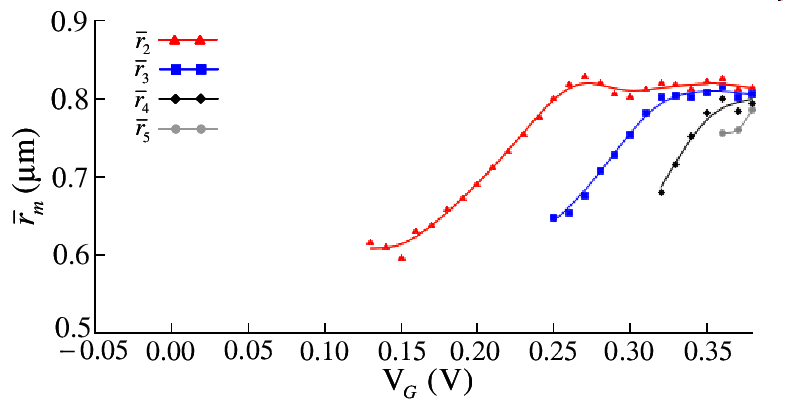}
\caption{(Color online) The mean nearest-neighbor distance $\bar{r}_{2}$, $\bar{r}_{3}$, $\bar{r}_{4}$, and $\bar{r}_{5}$ in the $x$ direction among the electrons organizing the two-, the three-, the four-, and the five-electron current, respectively.}
\label{fr4}
\end{figure}
%%%%%%%%%%%%%%%%%%%%%%%%%%%%%%%%%%

As a possible mechanism for the smoothing, we can indicate the observed temporal variation of $N_{\rm{G}}$ in our simulations.
To clarify this mechanism, we calculate the mean nearest-neighbor distance
\begin{eqnarray}
\bar{r}_{m}(V_{\rm{G}})=\left\langle \theta \left[\frac{3}{2}\bar{r}-r_{ij}(t)\right]r_{ij}(t) \bigg\vert \left(S_{\rm{CL}}\right)_{2}^{m}\right\rangle ,
\label{rm}
\end{eqnarray}
among the electrons forming the $m$-electron current, which is the current when $m$ electrons pass side by side through the constriction [see Eq. (\ref{sa2})].
Here, the region $S_{\rm{CL}}=\{(x,y) \vert \vert y\vert \leq 0.05 \ \mu$m $\}$ is set in order to measure the separations in the confined direction.
It is also to be noted that the average of $r_{ij}$ in Eq. (\ref{rm}) is limited to the middle distance of the nearest-neighbor and the second-neighbor distance.
From $\bar{r}_{m}$ for $m=2$, $3$, $4$, and $5$ plotted in Fig. \ref{fr4}, we find that the two-, three-, four-, and five-electron currents appear only in more than $V_{\rm{G}}=0.13$, $0.25$, $0.32$, and $0.36$ V, respectively.
These $V_{\rm{G}}$ are lower than the $V_{\rm{G}}$ at which the $m$ lines are observed in $n(\mathbf{r})$ [see Fig \ref{fr1}(a)].
In addition, we find intermediate states in which the $m$-electron current with long separations and the $(m+1)$-electron current with short separations coexist.
Since $G$ in the intermediate states takes between $m$- and $(m+1)$-electron current conductance, the effect can lead to the gradual change in $G$.
These results also reveal that the separations between electrons to pass side by side through the gate have a fluctuation margin about $0.1 \ \mu$m from the average interelectron separation $\bar{r}=2/\sqrt{\pi \bar{n}}=0.741 \ \mu$m.
A major factor in the temporally fluctuational separations will be specified in Sec. \ref{sec:subsec41}.

\section{\label{sec:sec4}Discussion}

In this section, we first discuss the origin of the current under the existence of the potential barrier and the mechanism smoothing the steps in the conductance.
Next, we compare the conductance in our system with the observed conductance in the device \cite{Rees}.
Finally, the obtained conductance from the LD simulations is briefly compared with that from the MD simulations.

\subsection{\label{sec:subsec41}Density-fluctuation-affected transport properties}

In this section, we propose two mechanisms of smoothing the steps in $G$.
We attribute the current for $V_{\rm{G}}<V_{\rm{G}}(1)$ to the intermittent disappearances of the barrier due to temporal fluctuation of the potential at both $\mathbf{r}_{\rm{G}}$ and $S_{\rm{F}}$.
This mainly results from the following two reasons.
First, the standard deviation of temporally variational potential energy in the channel, $\bar{\sigma}= \int_{\vert y\vert<D} d^2\mathbf{r}f(\mathbf{r};\vert y\vert<D)\sigma(\mathbf{r}) = 0.74(\pm 0.03)$ meV with Eq. (\ref{epf}), is much larger than $k_{B}T_{\rm{K}}$.
Second, the velocity distribution in the $y$ direction near the gate for $V_{\rm{G}} < V_{\rm{G}}(1)$ has the nearly zero mean value ($\sim 10^{-4}-10^{-7}$ meV in terms of the kinetic energy), and thus does not deviate from the equilibrium distribution.
Therefore, the conductor-insulator transition in our system is not caused by pinning and depinning of electrons at the constriction \cite{int6}.

From the conservation of energy, it is clear that electrons can pass through the gate only when the potential energy exceeds the barrier.
Therefore, $G$ up to $V_{\rm{G}}(1)$ may be approximately represented as $G\approx \tilde{G}=G(1)P(1)$, where $G(1)$ is the conductance at $V_{\rm{G}}(1)$ and $P(1)$ is the probability that the potential energy of the front electron exceeds the potential energy at $\mathbf{r}_{\rm{G}}$.
From the direct calculations in terms of the MD (see Supplemental Material) \cite{SM}, we verify that the temporal change of $\phi (\mathbf{r},t)$ almost satisfies the normal distributions as follows:
\begin{eqnarray}
\left\langle \delta \left[\epsilon-\phi (\mathbf{r},t)\right] \left\vert \left(\delta\mathbf{r}\right)_{1} \right.\right\rangle \approx \frac{\exp\left\{-\frac{\left[\epsilon-\Phi(\mathbf{r})\right]^2}{2\sigma(\mathbf{r})^2}\right\}}{\sqrt{2\pi \sigma(\mathbf{r})^{2}}},
\label{pd}
\end{eqnarray}
as in the case of the fluctuating electric field \cite{s10}.
Furthermore, it is probable that the temporal changes of $\phi (\mathbf{r},t)$ are uncorrelated between the two points of space.
This is because the potential of the electron just in front of the barrier is fluctuated mostly by two electrons in the rear [see Figs. \ref{f7}(a) and \ref{f6}(a)]; on the other hand, the potential at $\mathbf{r}_{\rm{G}}$ is fluctuated mostly by the electron of the other side across the barrier.
Hence, $\tilde{G}$ can be estimated as
\begin{eqnarray}
\tilde{G}&\approx &\frac{G(1)}{C}\int_{-\infty}^{\infty}d\epsilon \int_{\epsilon}^{\infty}d\epsilon ' \frac{\exp\left[-\frac{\epsilon'^2}{2\sigma_{\rm{F}}^{2}}\right] }{\sqrt{2\pi \sigma_{\rm{F}}^{2}}} \frac{\exp\left[-\frac{(\epsilon-E_{\rm{B}})^2}{2\sigma_{\rm{G}}^{2}}\right] }{\sqrt{2\pi \sigma_{\rm{G}}^{2}}}
\nonumber \\
&=&G(1) \ \mathrm{erfc}\left [\frac{E_{\rm{B}}}{\sqrt{2(\sigma_{\rm{F}}^{2}+\sigma_{\rm{G}}^{2})}} \right ],
\label{e5}
\end{eqnarray}
where
\begin{eqnarray}
C= \left.\mathrm{erfc}\left [\frac{E_{\rm{B}}}{\sqrt{2(\sigma_{\rm{F}}^{2}+\sigma_{\rm{G}}^{2})}} \right ]\right\vert_{E_{\rm{B}}=0},
\nonumber
\end{eqnarray}
is the probability for $E_{\rm{B}}=0$; $\sigma_{\rm{G}}(V_{\rm{G}})=\sigma(\mathbf{r}_{\rm{G}})$ and $\sigma_{\rm{F}}(V_{\rm{G}})= \int_{S_{\rm{F}}} d^2\mathbf{r}f(\mathbf{r};S_{\rm{F}})\sigma(\mathbf{r})$ are, respectively, the standard deviation of the potential energy at $\mathbf{r}_{\rm{G}}$ and in $S_{\rm{F}}$ (see Fig. \ref{fr2}); and erfc[$x$] is the complementary error function [solid line of Fig. \ref{fr1}(a)].
Although $\tilde{G}$ is estimated on the basis of only the one necessary condition, nevertheless $\tilde{G}$ is in good agreement with $G$ calculated in our simulations.

It should also be noted that the rising of $G$ takes place at $V_{\rm{G}}$ at which $E_{\rm{B}}$, $\sigma_{\rm{F}}$, and $\sigma_{\rm{G}}$ satisfy the relation 
\begin{eqnarray}
E_{\rm{B}}\simeq 2\sigma_{\rm{F}}+2\sigma_{\rm{G}},
\label{eb}
\end{eqnarray}
where tails of the potential distribution at the two different points touch just each other.
Therefore, this may give a method of an approximate estimation of the potential fluctuation from the potential energy.

%%%%%%%%%%%%%%%%%%%%%%%%%%%%%%%%%%
\begin{figure}[!t]
\center\includegraphics[width=86mm,clip]{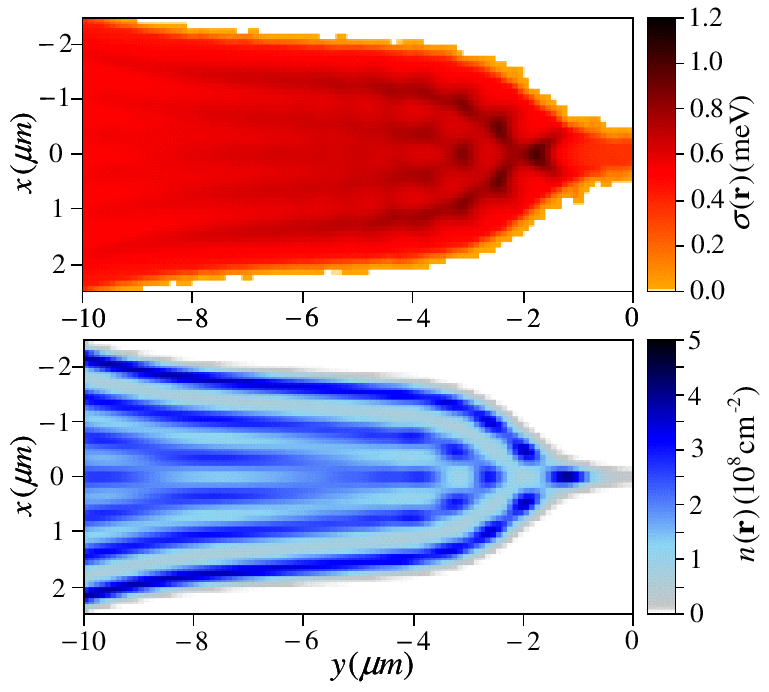}
\caption{(Color online) The spatial distribution of potential fluctuation $\sigma (\mathbf{r})$ (upper figure), and the density distribution $n(\mathbf{r})$ (lower figure) at $V_{\rm{G}}=0.05$ V in the region of $-10 \ \mu$m $\leq y \leq 0 \ \mu$m.
}
\label{fr5}
\end{figure}
%%%%%%%%%%%%%%%%%%%%%%%%%%%%%%%%%%

Concerning the choice of $y_{\rm{F}}$ in $S_{\rm{F}}$ [see Eq. (\ref{e4})], the discussion on $\tilde{G}$ holds well for the range of $-2.3 \ \mu$m $\leq y_{\rm{F}}\leq$ $-1.3 \ \mu$m in which the deviation $\vert \tilde{G}-G\vert$ at each $V_{\rm{G}}$ is less than $0.04$ M$\Omega^{-1}$.
Outside the range, $\tilde{G}$ becomes discrepant from $G$ because of large spatial variation in $\sigma(\mathbf{r})$.
Figure \ref{fr5} illustrates that $\sigma(\mathbf{r})$ increases in incommensurate regions in which two structures with the different number of the lines in $n(\mathbf{r})$ are frustrated \cite{int4,int5,sim2,i15}.

As seen in Sec. \ref{sec:sec3}, the temporal variation of $N_{\rm{G}}$ arises from the fluctuational interelectron separations in the confined direction.
The variational separations can be understood from an effect of the density fluctuation.
The root-mean-square displacement $\delta_{x}$ in the $x$ direction estimated from the harmonic approximation by equalizing $\delta_{x}^{2}\nabla^{2}\phi_{\rm{I}}(\bar{r})$ to $k_{\rm{B}}T_{\rm{K}}/2$ \cite{s9,s10}, i.e.,
\begin{eqnarray}
\delta_{x}=\sqrt{\frac{k_{\rm{B}}T_{\rm{K}}}{2\nabla^{2}\phi_{\rm{I}}(\bar{r})}} .
\label{er3}
\end{eqnarray}
is given by $\delta_{x}=0.119 \ \mu$m, which is comparable with the fluctuation margin.
The observed crystal-like ordering in the confined direction in the density patterns in Fig. \ref{f6} also supports of our estimation.
Therefore, we attribute grounds for the temporal change of $N_{\rm{G}}$ to the vibrational behavior.

In the above discussion, we suppose that the smooth rising from the insulating state to the first step in $G$ is due to the potential fluctuation, and the smooth steps in $G$ are caused by the vibration in the confined direction.
These dynamics can be commonly attributed by the density fluctuations.
Therefore, the magnitude of the density fluctuation seems to determine whether the steps in $G$ can be observed.

\subsection{\label{sec:subsec42}Comparison of our result with the observed conductance}

%%%%%%%%%%%%%%%%%%%%%%%%%%%%%%%%%%
\begin{figure}[!h]
\center\includegraphics[width=86mm,clip]{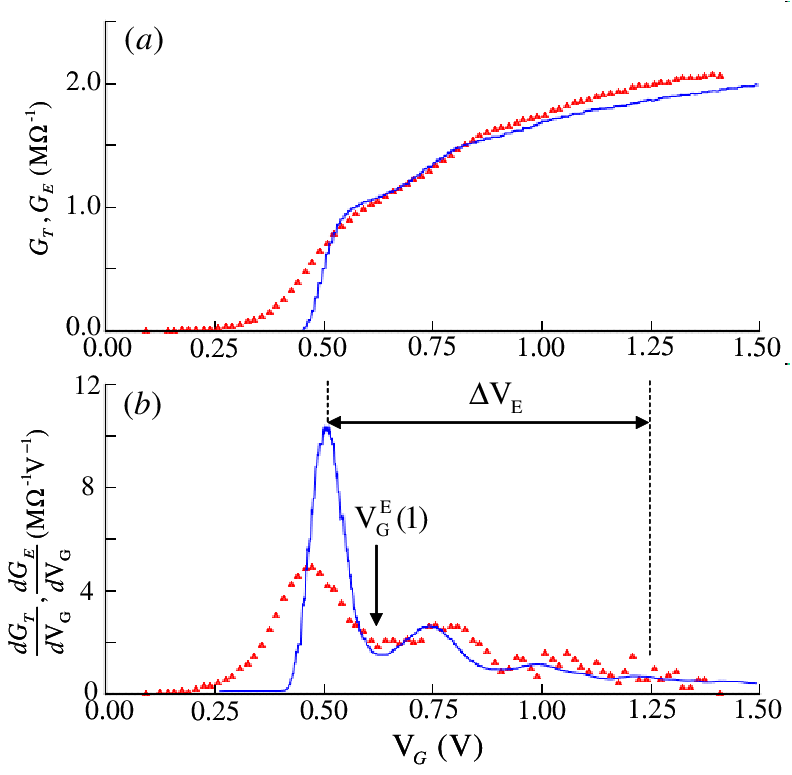}
\caption{ (Color online) (a) The measured conductance $G_{\rm{E}}$ in the experiment (blue solid line) \cite{Rees} and $G(V_{\rm{G}})$ in Fig. \ref{fr1}(a) transformed into $G_{\rm{T}}(V_{\rm{G}})=G( [ V_{\rm{G}}+V_{\rm{G}}(1) ] \Delta V_{\rm{S}}/\Delta V_{\rm{E}}-V_{\rm{G}}^{\rm{E}}(1))/g$ (red triangles) vs the gate voltage $V_{\rm{G}}$.
Here, $V_{\rm{G}}(1)$ and $\Delta V_{\rm{S}}$ are given in Fig. \ref{fr1}, and $\Delta V_{\rm{E}}$ and $V_{\rm{G}}^{\rm{E}}(1)$ are shown below.
(b) $dG_{\rm{E}}/dV_{\rm{G}}$ (blue solid line) \cite{Rees} and $dG_{\rm{T}}/dV_{\rm{G}}$ (red solid triangles), with respect to $V_{\rm{G}}$.
Here, the range $\Delta V_{\rm{E}}= 0.750$ V between the vertical dashed lines runs from the first to the fourth peak in $dG_{\rm{E}}/dV_{\rm{G}}$.
$V_{\rm{G}}^{\rm{E}}(1)=0.63$ V indicates $V_{\rm{G}}$, corresponding to the experimental counterpart of $V_{\rm{G}}(1)$ in the simulations.
}
\label{f12}
\end{figure}
%%%%%%%%%%%%%%%%%%%%%%%%%%%%%%%%%%

In this section, we illustrate the qualitative consistency between our and experiment conductance in order to stress the suitability of our approach for the electron system in the device \cite{Rees}.

The scaled and shifted $G$ are similar to the measured conductance $G_{\rm{E}}$ in Ref. \cite{Rees} as seen in Fig. \ref{f12}.
First, we select a scale factor $g=1.414$ so that the magnitude of $G$ at the first step fits into that of $G_{\rm{E}}$.
We confirm that the magnitude of $G$ is dependent on $\mu_{\rm{R}}^{1}$ or electron density as with the experimental results \cite{Rees}.

Second, we select a constant $\Delta V_{\rm{E}}/\Delta V_{\rm{S}}=3.3$ as a scale factor for $G$ in the $V_{\rm{G}}$ direction, where $\Delta V_{\rm{S}}$ and $\Delta V_{\rm{E}}$ are the amount of increase in $V_{\rm{G}}$ with an increment of $N_{\rm{G}}$ from $1$ to $4$ in the simulation and the experiment \cite{Rees}, respectively [see Figs. \ref{fr1}(b) and \ref{f12}(b)].
The choice of the scale factor stands on the consensus that a steplike increase in conductance is determined by $N_{\rm{G}}$.
Because $\Delta V_{\rm{E}}$ decreases with decreasing electron density, as observed in Ref. \cite{Rees}, and the density $\bar{n}=2.32\times 10^{8}$ cm$^{-2}$ in the simulations is lower than the experimental density \cite{Rees} $\bar{n}_{\rm{E}}=1.5\times 10^{9}$ cm$^{-2}$, the result of $\Delta V_{\rm{S}}<\Delta V_{\rm{E}}$ is reasonable.

Third, we shift $G$ so that $V_{\rm{G}}(1)$ agrees with the first-minimum-gate voltage $V_{\rm{G}}^{\rm{E}}(1)$ which corresponds to the minimum between the first and the second peak in $dG_{\rm{E}}/dV_{\rm{G}}$ [see Fig. \ref{f12}(b)].
Because the guard voltage $0.62$ V in the device corresponds to $0$ V in our system, $V_{\rm{G}}(1)$ is practically larger than $V_{\rm{G}}^{\rm{E}}(1)$ by $\{ V_{\rm{G}}^{\rm{E}}(1)-0.62\}-\{ V_{\rm{G}}(1)-0\} =0.135$ V.
The shift is also valid.
Indeed, the shift of conductance into the lower $V_{\rm{G}}$ direction is observed as electron density increases \cite{Rees}.

As a result of the above discussion, $G$ transformed into $G_{\rm{T}}$ in Fig. \ref{f12}(a) is almost in agreement with $G_{\rm{E}}$ for $V_{\rm{G}}>V_{\rm{G}}^{\rm{E}}(1)$, but is not good for $V_{\rm{G}}<V_{\rm{G}}^{\rm{E}}(1)$.
The disagreement for $V_{\rm{G}}<V_{\rm{G}}^{\rm{E}}(1)$ is also reasonable, for the amount of increase in $V_{\rm{G}}$ with the growth in conductance from the threshold of current flow to the first step is almost invariant with respect to electron density \cite{Rees}.
For $V_{\rm{G}}<V_{\rm{G}}(1)$, the shifted $G$ without scaling of $\Delta V_{\rm{E}}/\Delta V_{\rm{S}}$ is actually coincident with $G_{\rm{E}}$ for $V_{\rm{G}}<V_{\rm{G}}^{\rm{E}}(1)$.

\subsection{\label{sec:subsec43}Langevin dynamics simulation}

We also calculate the conductance in the molecular dynamics under the Langevin thermostat, which is known as the Langevin dynamics, to check how the results depend on our choice of the No\'se-Hoover thermostat.
The detailed description on the method is given in Appendix \ref{sec:a3}.

Figure \ref{ldc} is the corresponding plot to Fig. \ref{fr1}(a) for the LD simulations.
The obtained $G_{\rm{L}}$ in Fig. \ref{ldc} under the averages of $4,300,000$ time steps and $16$ different initial conditions reproduces the steplike conductance as observed in Ref. \cite{Rees}, and shows qualitatively similar behavior to $G$ in Fig. \ref{fr1}(a).
However, we can find the sharper rising to the first step of $G_{\rm{L}}$ than that of $G$.
This deference reflects on the observed smaller potential fluctuation in the LD simulations than that in the MD simulations (see Sec. \ref{sec:subsec41}).
Because the characteristic time scale of the relaxation of electron motion is $1/\xi$, which is introduced in Eq. (C1) \cite{WU}, the reduction in potential fluctuation may result from the damping.
A more detailed comparison between MD and LD will be discussed in future work.

%%%%%%%%%%%%%%%%%%%%%%%%%%%%%%%%%%
\begin{figure}[!h]
\center\includegraphics[width=86mm,clip]{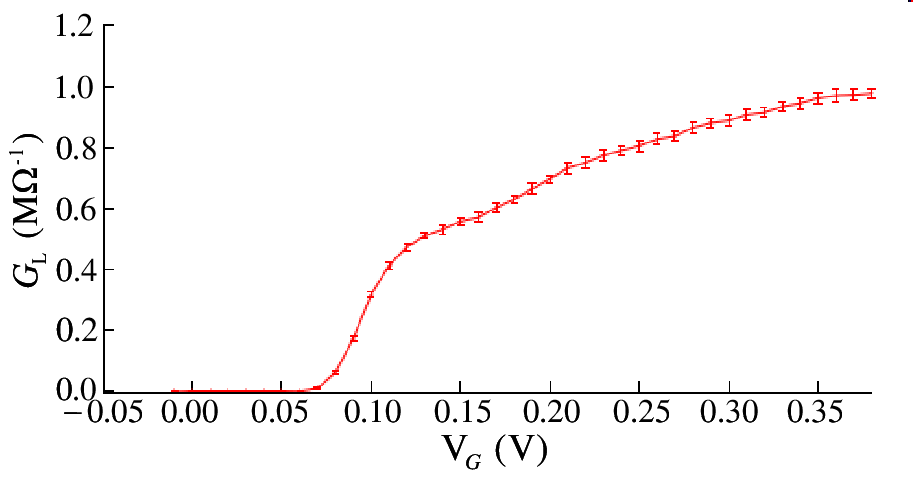}

\caption{(Color online) (a) The dc conductance $G_{\rm{L}}$ obtained from the LD simulations versus the gate voltage $V_{\rm{G}}$, where each error bar represents the standard deviation of the ensemble average for each $G_{\rm{L}}$.
}
\label{ldc}
\end{figure}
%%%%%%%%%%%%%%%%%%%%%%%%%%%%%%%%%%

\section{\label{sec:sec5}Conclusion}

In conclusion, we reproduce the steplike conductance in a classical 2D electron system confined in the point-contact device on liquid $^{4}$He by the molecular dynamics simulation.
Conductance in two-dimensional electrons confined in the shape of a microchannel with a point contact by electric fields is calculated as a function of the confinement strength of the contact.
It is confirmed that the number of the steps corresponds to the number of stream lines formed by the flow of self-organizing distributed electrons at the contact.
This result supports the expectation by Rees {\it et al}. \cite{Rees} in which the conductance does not originate from quantum effects but from classical many-particle effects at the contact.
We verify that a potential barrier exists in the contact for the stronger confinement than that at the first step, and the barrier disappears for the weaker confinement in which the constriction width of the contact increases with weakening confinement.
In the strong confinement, the rising of the conductance can be attributed to intermittent disappearances of the barrier due to temporal fluctuation of the electrostatic potential.
In the weak confinement, the number of electrons to pass simultaneously through the contact is incremented through the intermediate state in which the number varies temporally.
Based on these results, we suppose that the density fluctuation smooths the steps in the conductance.

{\it Note added}. Recently, a similar paper on the same subject based on both the experiment and MD has been published \cite{Rees2},
Although our work differs in temperature range from the paper, which makes a study of the transport of the Wigner crystal \cite{Rees2}, the electron dynamics clarified by our investigations also provides information to help them understand the mechanism.

\begin{acknowledgments}

The authors thank D. Rees and K. Kono for fruitful discussions.
This work was partially supported by the Ministry of Education, Culture, Science and Technology (MEXT), Japan (Grant No. 21540384), and by the Global COE program, The Next Generation of Physics, Spun from Universality and Emergence, from MEXT Japan. The author also thanks the Yukawa International Program for Quark-Hadron Sciences at Yukawa Institute for Theoretical Physics, Kyoto University.
The numerical calculations were carried out on the SR16000 and Altix3700 BX2 at YITP in Kyoto University.

\end{acknowledgments}

\appendix

\section{\label{sec:a1}Estimation of resistance due to the electron-helium gas atom and the electron-ripplon scatterings}

In this appendix, we investigate the transport process in the channel, as in Ref. \cite{Rees}.
It is demonstrated that the electron-helium gas atom and the electron-ripplon scatterings are ineffective in the experimental resistance $R_{\rm{E}}> 5\times 10^{-1}$ M$\Omega$ for the system.

For the bulk 2D systems, the effective collision frequencies, which are defined by the relaxation time in the Drude formula, 
are given in Ref. \cite{cc4}.
For the setting in Ref. \cite{Rees}: $T=1.2$ K, $z_{\rm{He}}=1.5 \ \mu$m, $F_{\perp}=6.5\times 10^{2}$ V$/$cm, the values of the frequencies are $\nu_{\rm{g}}=5.6\times 10^{9}$ s$^{-1}$ for the the electron-helium gas atom scattering, and $\nu_{\rm{r}}=1.4\times 10^{9}$ s$^{-1}$ for the electron-ripplon scattering under the single-electron approximation.
We also confirm that the predominance of $\nu_{\rm{g}}$ upon the resistance is unchanged even if the electron-electron scattering effect on $\nu_{\rm{r}}$ is considered under the complete control approximation.
Since these frequencies make the mean free path to be of the order of $0.1 \ \mu$m, the system is diffusive \cite{in4,in5,in6,in7,in8,in9,in10}.

For the transport in diffusive regions, the macroscopic geometrical effects as in the Maxwell resistance are dominant \cite{in10}.
Based on the Poisson equation and the Ohm law, the geometrical effect of the bottleneck [$10\ \mu$m $\leq \vert y \vert \leq 15 \ \mu$m; see Fig. \ref{f2}(a) and Ref. \cite{Rees}] upon the resistance is calculated in Ref. \cite{r2-2} as
\begin{eqnarray}
R_{\rm{B}}&=&\frac{\rho^{0}}{\pi}\left\{ \frac{ W_{\rm{n}}^{2}+W_{\rm{w}}^{2}}{W_{\rm{n}}W_{\rm{w}}}\tanh^{-1}\left[\frac{W_{\rm{n}}}{W_{\rm{w}}} \right] \right.
\nonumber \\
&+&\left. \frac{ W_{\rm{w}}^{2}-W_{\rm{n}}^{2}}{W_{\rm{n}}W_{\rm{w}}}\tan^{-1}\left[\frac{W_{\rm{n}}}{W_{\rm{w}}} \right]+\log \left[\frac{W_{\rm{w}}^{4}-W_{\rm{n}}^{4}}{8W_{\rm{i}}^{2}W_{\rm{w}}^{2}}\right]\right\},
\nonumber \\
\label{br}
\end{eqnarray}
where $\rho^{0}=m(\nu_{\rm{g}}+\nu_{\rm{r}})/\bar{n}_{\rm{E}}e^2$ is the bulk resistivity with the reported density $\bar{n}_{\rm{E}}=1.5\times 10^{9}$ cm$^{-2}$ in Ref. \cite{Rees}, and $W_{\rm{n}}=10 \ \mu$m and $W_{\rm{w}}=2W_{\rm{n}}$ are the widths at the narrow and the wide section of the bottleneck, respectively.
Therefore, the resistance in the channel is given by $R=\rho^{0}L/W_{\rm{n}}+2R_{\rm{B}}=4.7\times 10^{-2}$ M$\Omega$, where $L=20 \ \mu$m is the channel length without the bottleneck portion.
Note that $R$ is $10$\% of $R_{\rm{E}}$.

In addition, we take into account the boundary scattering effects to the resistivity \cite{in1}.
In the boundary scattering theory \cite{in18,in19,in20,in21,in22,in23}, the steady-state Boltzmann equation on the coordinate system in Fig. \ref{f2}(a) under the free-electron approximation is given by
\begin{equation}
\mathbf{v} \cdot \frac{\partial \Psi_{1}(\mathbf{r},\mathbf{v})}{\partial \mathbf{r}}-\frac{e F_{\parallel}}{m}\frac{\partial \Psi_{0}(v)}{\partial v_{y}}=-\frac{\Psi_{1}(\mathbf{r},\mathbf{v})}{\tau},
\label{be}
\end{equation}
which can be solved under an ideal hard-wall boundary condition, where $\Psi_{0}$ is the equilibrium electron distribution function to be the product of the constant density $\bar{n}_{\rm{E}}$ and the Maxwell-Boltzmann distribution function, $\Psi_{1}$ is a deviation of an electron distribution function from $\Psi_{0}$, $F_{\parallel}$ is an in-plane driving field, and $\tau$ is a relaxation time \cite{in19,in25,in26}.
In Eq. (\ref{be}), we simply adopt $\tau=1/(\nu_{\rm{g}}+\nu_{\rm{r}})$ because the elastic scattering is dominant. We also set the boundary at $x=W_{\rm{n}}/2$ and $x=-W_{\rm{n}}/2$ to simplify the argument.
For the diffusive boundary scattering \cite{in18,in19}, the resistivity $\rho$ is given by
\begin{eqnarray}
&\rho&=\rho^{0}/\left\{ 1-\frac{\tau}{W_{\rm{n}}}\sqrt{\frac{2k_{\rm{B}}T}{m\pi}}\right.
\nonumber \\
&\!\! + \!\!&\left.\frac{\tau}{W_{\rm{n}}}\sqrt{\frac{2k_{\rm{B}}T}{m\pi}}2 \! \int^{\infty}_{0} \!\!\!\!\!\! d v_{x} v_{x} \exp\left[\! - \!\left(v_{x}^{2}\! +\!\frac{\tau}{W_{\rm{n}}}\sqrt{\frac{m}{2k_{\rm{B}}T}}\frac{1}{v_{x}}\right)\right]\right\}.
\nonumber \\
\label{rho}
\end{eqnarray}
This result leads to $\rho=1.05\rho^{0}$, and thus, the effects little affect the channel resistance.
Therefore, the correlation effects between electrons in the external potential are dominant for reproducing the value of $R_{\rm{E}}$.

\section{\label{sec:a2}Calculation method of physical quantities}
The one-particle quantities we calculate are the density distribution function averaged over the $0.1$ $\mu$m square $\delta\mathbf{r}$ centered around the position $\mathbf{r}$:
\begin{eqnarray}
n(\mathbf{r})= \left\langle \sum_{i}\delta (\mathbf{r}-\mathbf{r}_{i}(t)) \right\rangle ,
\label{n}
\end{eqnarray}
the average potential energy of electrons in $\delta\mathbf{r}$,
\begin{eqnarray}
\Phi (\mathbf{r})=\left\langle \phi (\mathbf{r}_{i},t) \left\vert \left(\delta\mathbf{r}\right)_{1} \right. \right\rangle,
\label{ep}
\end{eqnarray}
and the standard deviation of temporal fluctuation of $\Phi (\mathbf{r})$,
\begin{eqnarray}
\sigma(\mathbf{r})=\left\{ \left\langle \left\vert \phi(\mathbf{r}_{i},t)-\Phi\left[\mathbf{r}_{i}(t)\right]\right\vert^{2} \Big\vert \left(\delta\mathbf{r}\right)_{1} \right\rangle \right\}^{\frac{1}{2}},
\label{epf}
\end{eqnarray}
where $\langle  \cdots \vert (\delta\mathbf{r})_{1} \rangle$ represents the ensemble average under a conditional time average for electrons in $\delta\mathbf{r}$, when electrons are present in $\delta\mathbf{r}$.
Here, $\delta \mathbf{r}$ is created by dividing the $xy$ plane into 0.1 $\mu$m square mesh, and $n$ in $(\delta\mathbf{r})_{n}$ denotes that the averaged quantity is the $n$-particle quantity.
The conditional time average is defined by
\begin{eqnarray}
\frac{1}{t_{\rm{P}}(\delta \mathbf{r})}\int_{0}^{t_{\rm{T}}}dt \int_{\delta\mathbf{r}}d^{2}\mathbf{r}'\sum_{i}\delta \left[\mathbf{r}'-\mathbf{r}_{i}(t)\right]\cdots ,
\label{cta}
\end{eqnarray}
where $t_{\rm{T}}$ is the calculation time, and
\begin{eqnarray}
t_{\rm{P}}(\delta \mathbf{r})\equiv \int_{0}^{t_{\rm{T}}}dt\int_{\delta\mathbf{r}}d^{2}\mathbf{r}''\sum_{k}\delta \left[ \mathbf{r}''-\mathbf{r}_{k}(t) \right]
\nonumber
\end{eqnarray}
is the total presence time of electrons in $\delta\mathbf{r}$.
We also confirm that the quantities are almost in agreement with those calculated with $\delta\mathbf{r}$ to be 0.05 $\mu$m squares.
In addition, we use another conditional average for a two-particle quantity as follows:
\begin{widetext}
\begin{eqnarray}
\langle  \cdots \vert (S)_{2}^{m} \rangle = \left\langle\frac{1}{t_{\rm{2P}}(S;m)}\int_{0}^{t_{\rm{T}}}dt \delta (m-N_{\rm{S}}(t)) \int_{S}d^{2}\mathbf{r}' \int_{S}d^{2}\mathbf{r}''\sum_{i>j}\delta \left[ \mathbf{r}'-\mathbf{r}_{i}(t) \right] \delta \left[ \mathbf{r}''-\mathbf{r}_{j}(t) \right] \cdots \right \rangle_{\rm{E}},
\label{sa2}
\end{eqnarray}
where $\langle \cdots \rangle_{\rm{E}}$ is the ensemble average, $N_{\rm{S}}(t)$ is the number of electrons in an area $S$ at time $t$:
\begin{eqnarray}
N_{\rm{S}}(t)=\int_{S}d^{2}\mathbf{r}'\sum_{k}\delta \left[\mathbf{r}'-\mathbf{r}_{k}(t)\right],
\nonumber
\end{eqnarray}
and $t_{\rm{2P}}(S;m)$ is the product of the number of pairs among $m$ electrons and the presence time when just $m$ electrons are present in $S$,
\begin{eqnarray}
t_{\rm{2P}}(S;m)=\int_{0}^{t_{\rm{T}}}dt \delta \left[m-N_{\rm{S}}(t)\right]\int_{S}d^{2}\mathbf{r}' \int_{S}d^{2}\mathbf{r}''\sum_{k>l}\delta \left[\mathbf{r}'-\mathbf{r}_{k}(t)\right] \delta \left[\mathbf{r}''-\mathbf{r}_{l}(t)\right].
\nonumber
\end{eqnarray}
\end{widetext}
Thus, $\langle  \cdots \vert (S)_{2}^{m} \rangle$ describes the ensemble and the conditional time averages, among $m$ electrons in $S$, when just $m$ electrons are present in $S$.
Incidentally, all of the introduced quantities in this section are time and ensemble averaged over 1,200,000 time steps and 32 different initial conditions, respectively.

\section{\label{sec:a3}Langevin dynamics simulation method}

In this appendix, we briefly explain the method of LD simulation to calculate the conductance of electrons \cite{AT}.
The equation of motion for the electrons in the LD simulations is given by
\begin{eqnarray}
m\frac{d^{2}r_{i}^{\alpha}}{dt^{2}}=F_{i}^{\alpha}-m \xi \dot{r_{i}}^{\alpha}+\tilde{F}_{i}^{\alpha},
\label{lds}
\end{eqnarray}
associated with Eqs. (\ref{e3-3}) and (\ref{e3-5}), where $F_{i}^{\alpha}$ consists of the first, third, and fourth terms of the right-hand side in Eq. (\ref{e3-1}), $\xi$ is the friction constant, and $\tilde{F}_{i}^{\alpha}$ is a random force, reproducing the thermal noise with zero mean value and variance $\langle \tilde{F}_{i}^{\alpha}(t)\tilde{F}_{j}^{\beta}(t') \rangle = 2mk_{\rm{B}}T\xi \delta_{ij}\delta_{\alpha \beta}\delta (t-t')$.
If we regard $\xi$ as the collision frequency through the Einstein' relation, we may estimate $\xi$ as $\xi=\nu_{\rm{g}}+\nu_{\rm{r}}=7.0\times 10^{9}$ s$^{-1}$ (see Appendix \ref{sec:a1} for $\nu_{\rm{g}}$ and $\nu_{\rm{r}}$).
For the value to be $\xi \ll 1/\tau_{\rm{s}}$, we have no reason to neglect the inertial term, where $1/\tau_{\rm{s}}$ is the characteristic frequency for the correlation between electrons.
To integrate Eq. (\ref{lds}), we adopt the Ermak's approach \cite{EB, AT}, which performs properly the stochastic integration \cite{C}.
This LD simulation is reduced to an overdamped LD simulation for $t_{0}\xi \gg 1$, and our MD simulation without the No\'se-Hoover thermostat in the limit $\xi \rightarrow 0$.
The other numerical methods and the simulation setup are unchanged from those of the MD simulations (see Sec. \ref{sec:sec2}).

\end{document}